\documentclass{vietnam}
\usepackage{amsmath}

\def\Journal#1#2#3#4{{#1} {\bf #2}, #3 (#4)}

\def\PRD{{\em Phys. Rev.} D}

\def\be{\begin{equation}}
\def\ee{\end{equation}}
\def\bea{\begin{eqnarray}}
\def\eea{\end{eqnarray}}

\newcommand{\Photo}{}

\begin{document}
\vspace*{4cm}
\title{Disentangling CMB $\mu$ and $y$ spectral distortions from foregrounds with poorly defined spectral shapes}

\author{Mihalchenko A.O.$^{1}$}
\address{{\tiny $^\textit{1}$}Astro-Space Center of P.N. Lebedev Physical Institute, Profsoyusnaya 84/32}

\maketitle\abstracts{
We have presented a new approach to separate small spectral $\mu$ and $y$ distortions of the CMB from foreground components with poorly defined spectral shapes. Our linear method, called the Least Response Method (LRM), is based on the idea of simultaneously minimizing the response to all possible foregrounds and photon noise while maintaining a constant response to the useful signal. We compared our approach with the mILC method, which is a modification of the Internal Linear Combination previously used for CMB anisotropy maps, and proved the advantages of LRM. In addition, we found the optimal temperature of the telescope optical system for any experiments related to the study of the CMB $\mu$ distortions.}

\section{Introduction}

The observed frequency spectrum of the CMB is perfectly approximated by a black body spectrum. 
A black body in thermal equilibrium emits radiation according to the Planck's law.
At a redshift $\sim 2 \times 10^6$ the processes that produce photons drop out of equilibrium and only Compton scattering is active. It can still redistribute CMB photons in energy but it does not change the number density. If energy injection happens after this point, we expect a spectral deviation from the Planck's law that leads to a Bose-Einstein distribution with a non-vanishing chemical potential. This deviation from a black body is called a $\mu$ distortion. The epoch of such distortions continues from $z \sim 2 \times 10^6$ to $z \sim 5 \times 10^4$. Much later in time, after reionization, when CMB photons pass through a hot electron gas, affected only by Compton scattering, another distortion of the spectrum is created -- a $y$ distortion. Both $\mu$ and $y$ distortions are very small, and their detection requires high precision. Yet it is complicated by the presence of foregrounds of various origin  (dust, cosmic infrared background, synchrotron radiation, free-free emission etc.) which are several orders of magnitude greater than the distortion signals and are poorly predictable. In addition, the optical system of the telescope generates noise and systematic errors. This means that efficient data processing methods are highly required.

\section{The Least Response Method and Moment Internal Linear Combination}

\begin{figure*}[!htbp]
\centerline{
\includegraphics[width=0.5\textwidth]{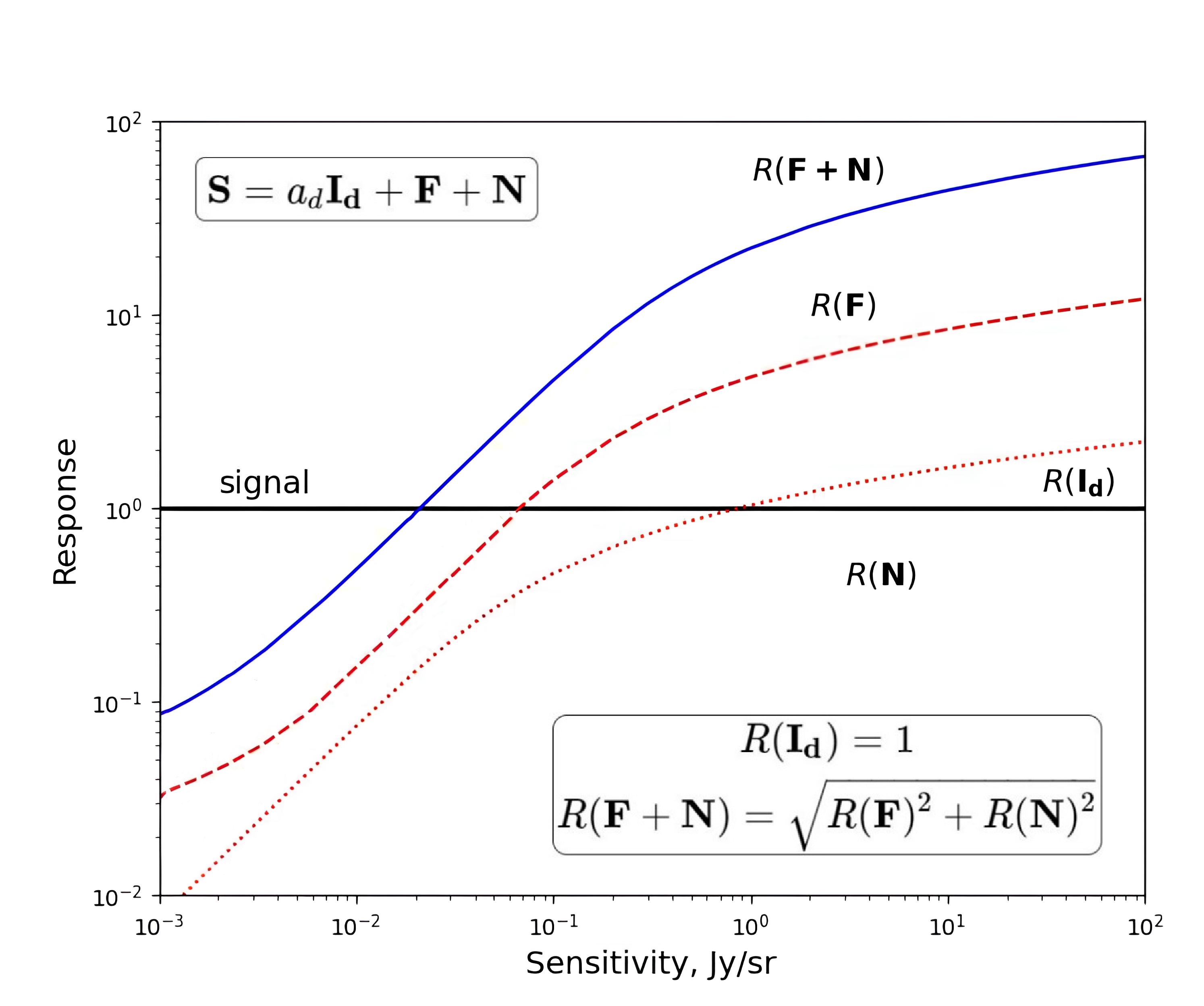}
\includegraphics[width=0.54\textwidth]{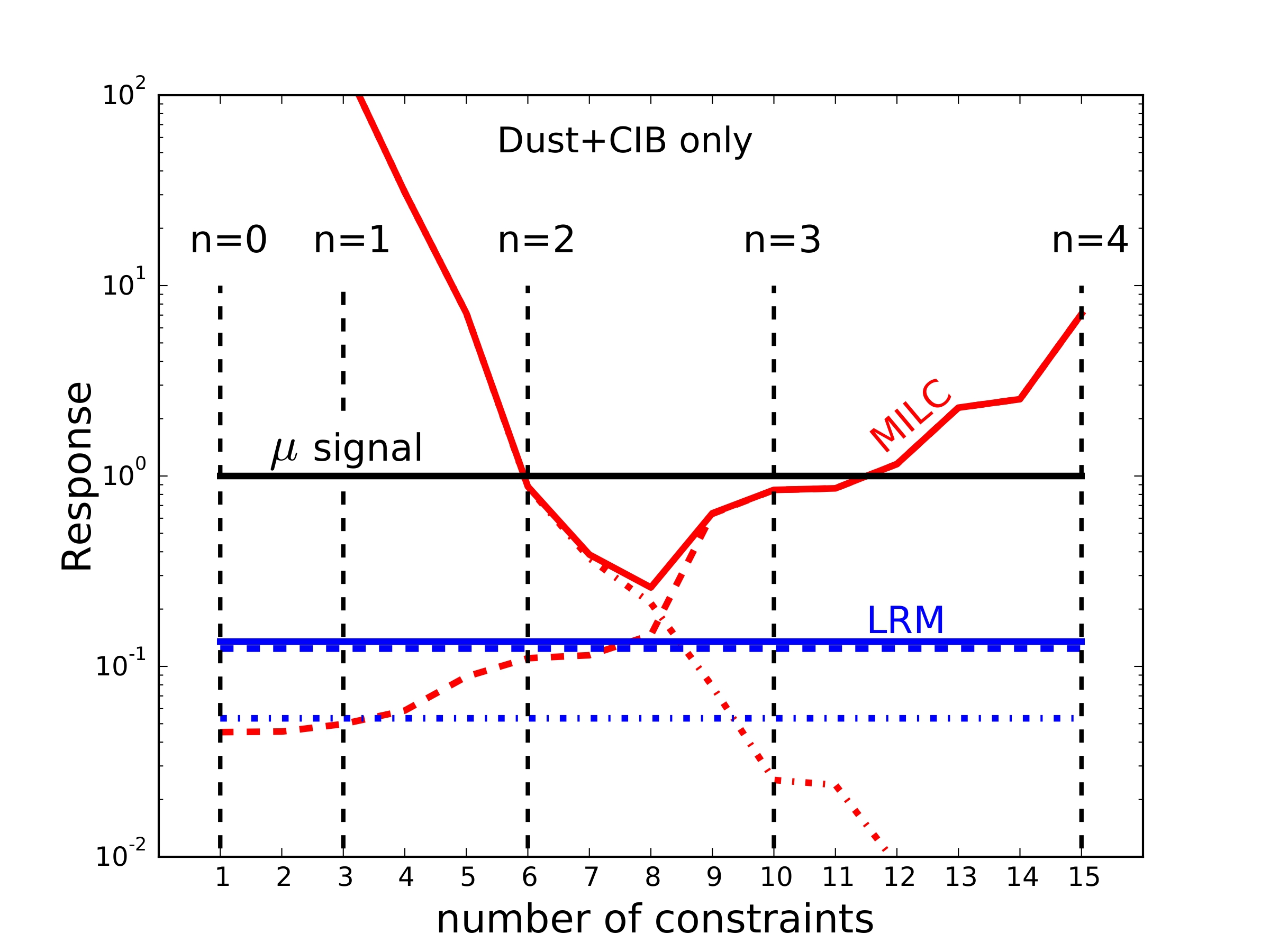}
}
\caption{{\it Left panel.} The evolution of the responses with respect to the sensitivity of the experiment. When sufficient sensitivity is reached, the response to the foregrounds becomes negligibly small compared to the response to the signal of interest. It's worth noting that the response to the distortion $R({\bf I_d})$ is constant and equal to one by definition. The goal of any reasonable approach is to drive the responses to foreground and noise under the constant horizontal line  (the response to the signal) for as high sensitivity values as possible in order to get a decent result even at mediocre levels of sensitivity.
{\it Right panel.} Red solid line: total response to foreground+noise for MILC as a function of the number of constraints.
Red dashed line: MILC response to noise. Red dash-dotted line: MILC response to foreground. The blue solid, dashed, and
dash-dotted lines: LRM responses to foreground+noise, noise, and foreground respectively. }
\label{pic:1}
\end{figure*}

To solve the aforementioned problems, we tested the methods that have been successfully used to analyze CMB anisotropy maps and process data from the WMAP and Planck experiments. Among these approaches are "blind" methods such as the \textit{Internal Linear Combination}~\cite{ry} (ILC), which only assumes the knowledge of the signal of interest SED and considers all other components of the observed signal as unmodeled noise, as well as hybrid methods ~\cite{sto,ch,ro} which combine elements of the blind approaches with partial use of information about foregrounds (by using constraints and moment expansion). Although these methods proved to be effective, they also have certain disadvantages. ILC is limited by the presence of large foreground components that have a nonzero projection onto the signal of interest, and given that the foregrounds are usually several orders of magnitude greater than the distortions, the result can get biased. As for MILC, a large number of constraints leads to a large response to unmodeled noise. Imposing too many strict conditions limits the number of degrees of freedom and increases the contribution of noise to the estimation of the signal of interest. 

\begin{figure*}[!htbp]
\centerline{
\includegraphics[width=0.5\textwidth]{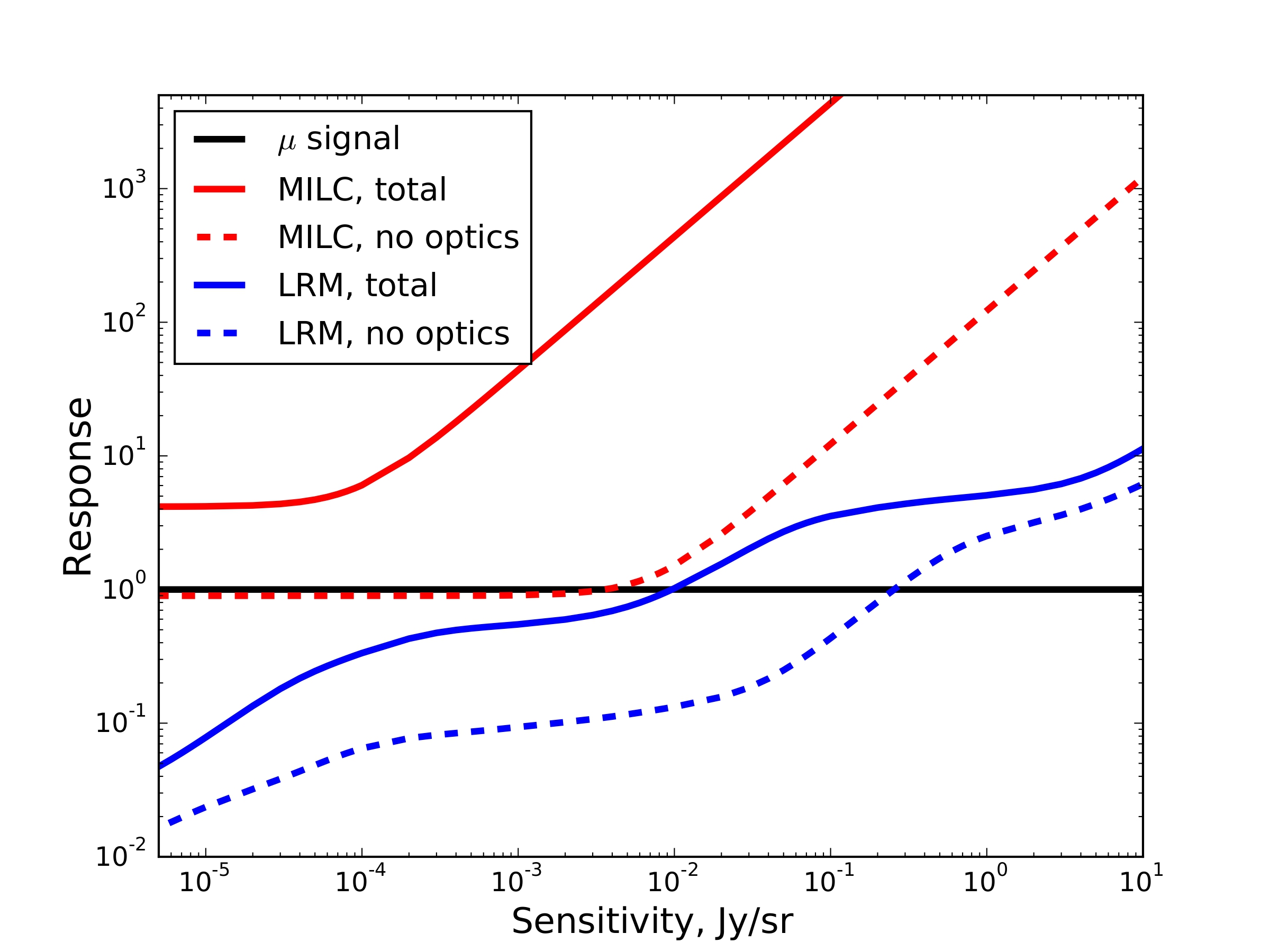}
\includegraphics[width=0.5\textwidth]{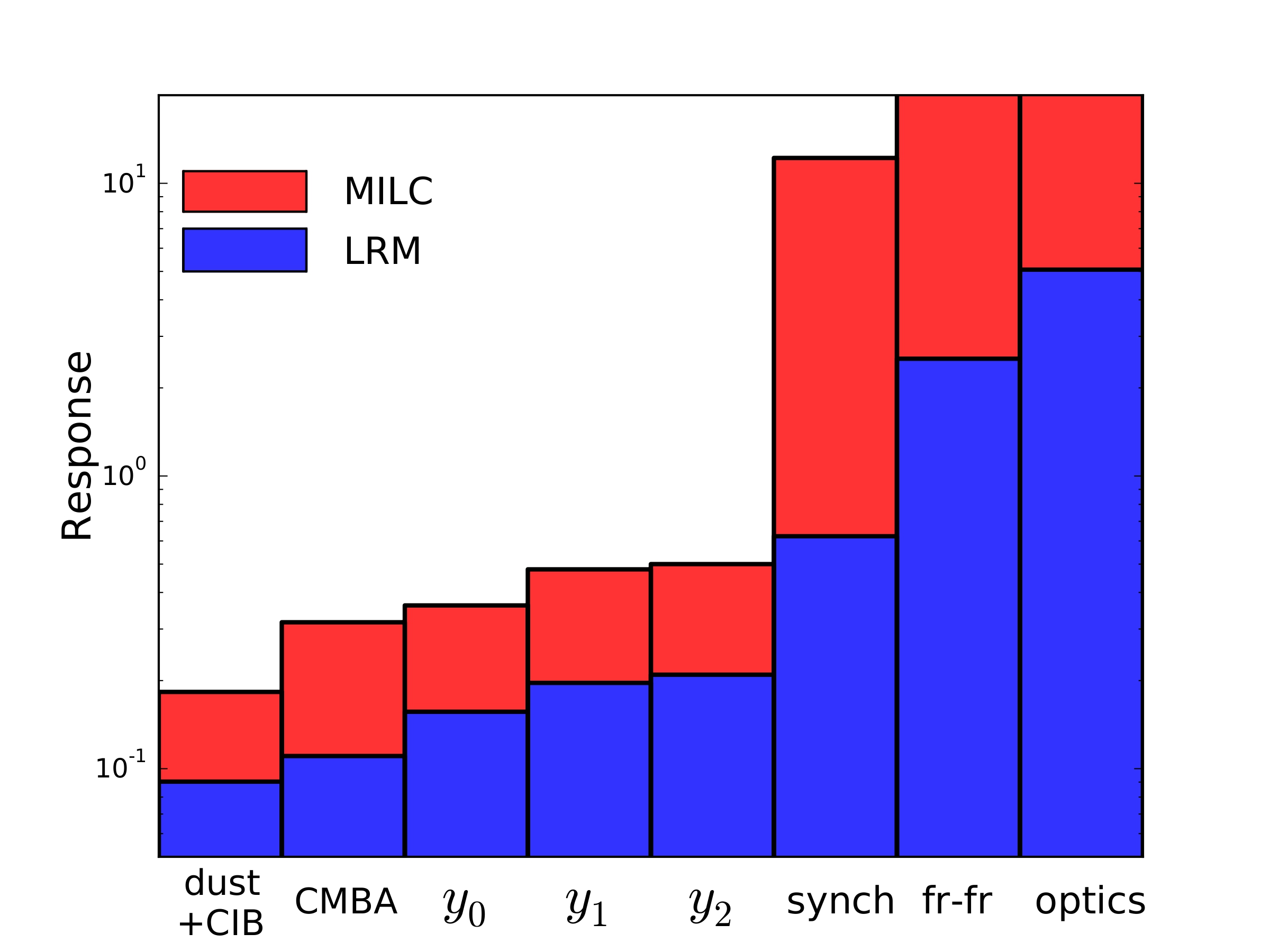}
}
\caption{ The result of applying the MILC and LRM methods to separate the $\mu$ signal. {\it Left panel.} The dependencies of the total noise+foreground response on the sensitivity of the experiment in the presence of all the main foregrounds are shown. The dashed lines correspond to an "ideal" experiment with no influence from the instrument optical system.
{\it Right panel.} The histogram represents the response to foreground+noise with the sequential addition (one by one from
left to right) of various components to the studied signal. The first column shows the response in the presence of only dust+CIB. The last column shows the response if all the components listed along the horizontal axis are taken into account.}
\label{pic:2}
\end{figure*}

In an attempt to tackle these problems we have devised an approach of our own –- the \textit{Least Response Method~\cite{no,ma}} (LRM). By applying this method, we have separated $\mu$ and $y$ spectral distortions from the signal inundated with foregrounds of cosmic (galactic dust, cosmic infrared background, synchrotron radiation, free-free emission) and instrumental (the optical system of the telescope) origin. The main problem, which has long been considered practically unsolvable, is the fact that the contribution to the total signal from foregrounds along the line of sight consists of a large number of sources with different spectral characteristics described by different spectral parameters. Even for a single line of sight, the observed cosmic foreground spectrum is actually a superposition
of spectra with different parameters

Any foreground with parameters from a certain pre-determined parameters domain is practically eliminated by the LRM. The response to the foreground with parameters lying in this domain becomes negligibly small compared to the response to the desired signal as long as the sensitivity of an experiment is sufficient. Note that the only assumption about the amplitude of a foreground signal in our approach is that the the amplitude cannot exceed a certain value estimated in advance from previous experiments.

We have shown that all of the previously used approaches in this case either do not work at all (the "blind" ILC method), or result in a large response to photon noise. This means that such methods can be used only when sensitivity of an experiment is high. The sensitivity of the upcoming experiments is limited, hence the need to use the most effective data processing method. The developed approach proves to be the one that can claim this role.

\begin{figure*}[!htbp]
\centerline{\includegraphics[width=0.8\textwidth]{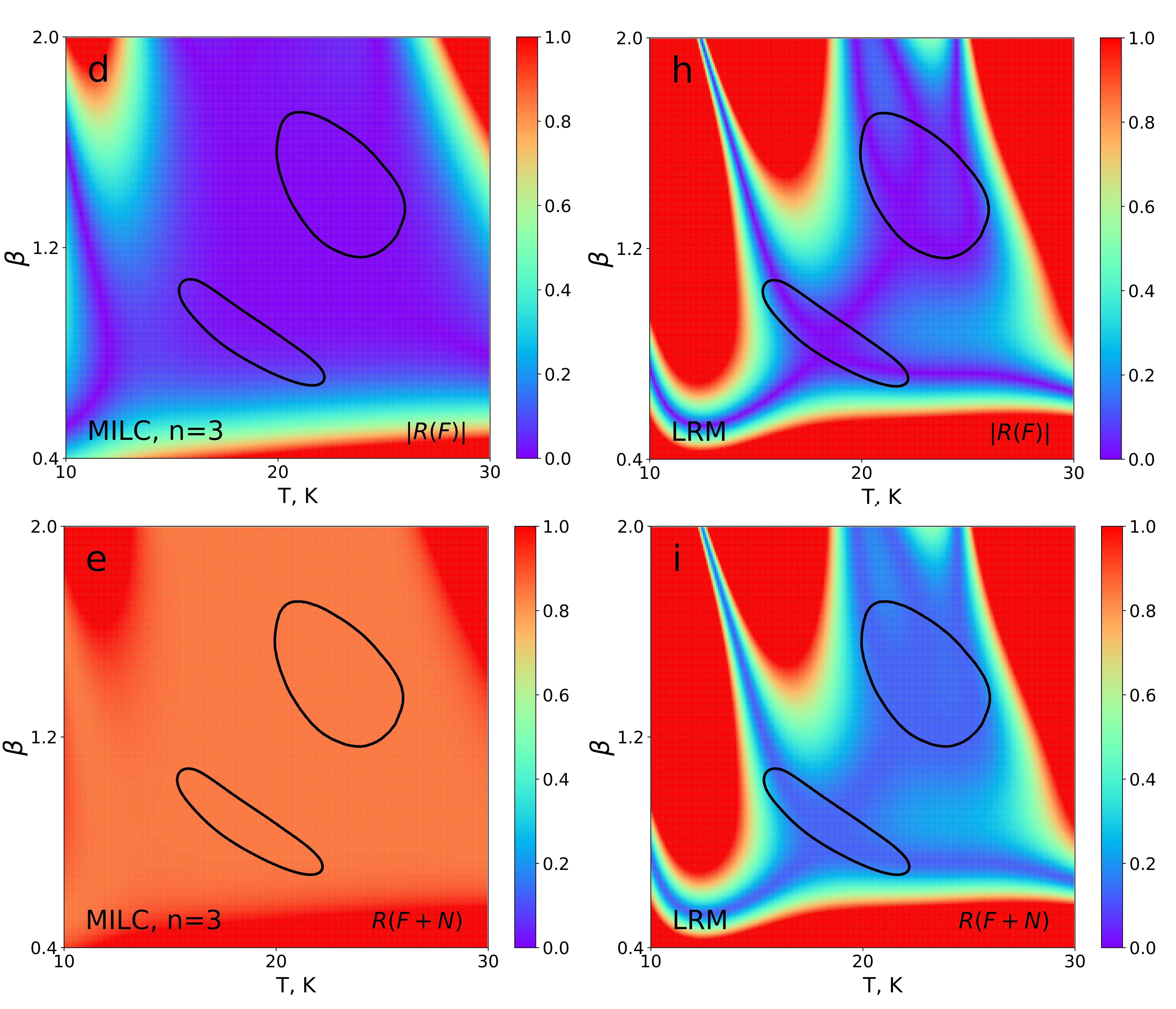}}
\caption{MILC and LRM methods for separating $\mu$
    distortions when only dust and CIB are assumed as foregrounds. Solid black contours mark the boundaries of the temperature and spectral index parameters domains (determined with the help of the Planck data). The left column depicts MILC responses to the foreground and noise with $10$ total constraints (this amounts to $n=3$ moments). The right column -- LRM responses. Dark red indicates the regions where the response to the foreground exceeds the response to the signal of interest.
}
\label{pic:3}
\end{figure*}

Moreover, modeling the spectrum emitted by the telescope optics is an extremely difficult enterprise. The characteristics of the mirror elements change during an observation in an unpredictable way. However, the developed algorithm allows us to overcome these issues. Only three quantities are necessary: the minimum and maximum possible temperatures of the mirror surface, and its maximum possible emissivity.

\begin{figure*}[!htbp]
\centerline{
\includegraphics[width=0.54\textwidth]{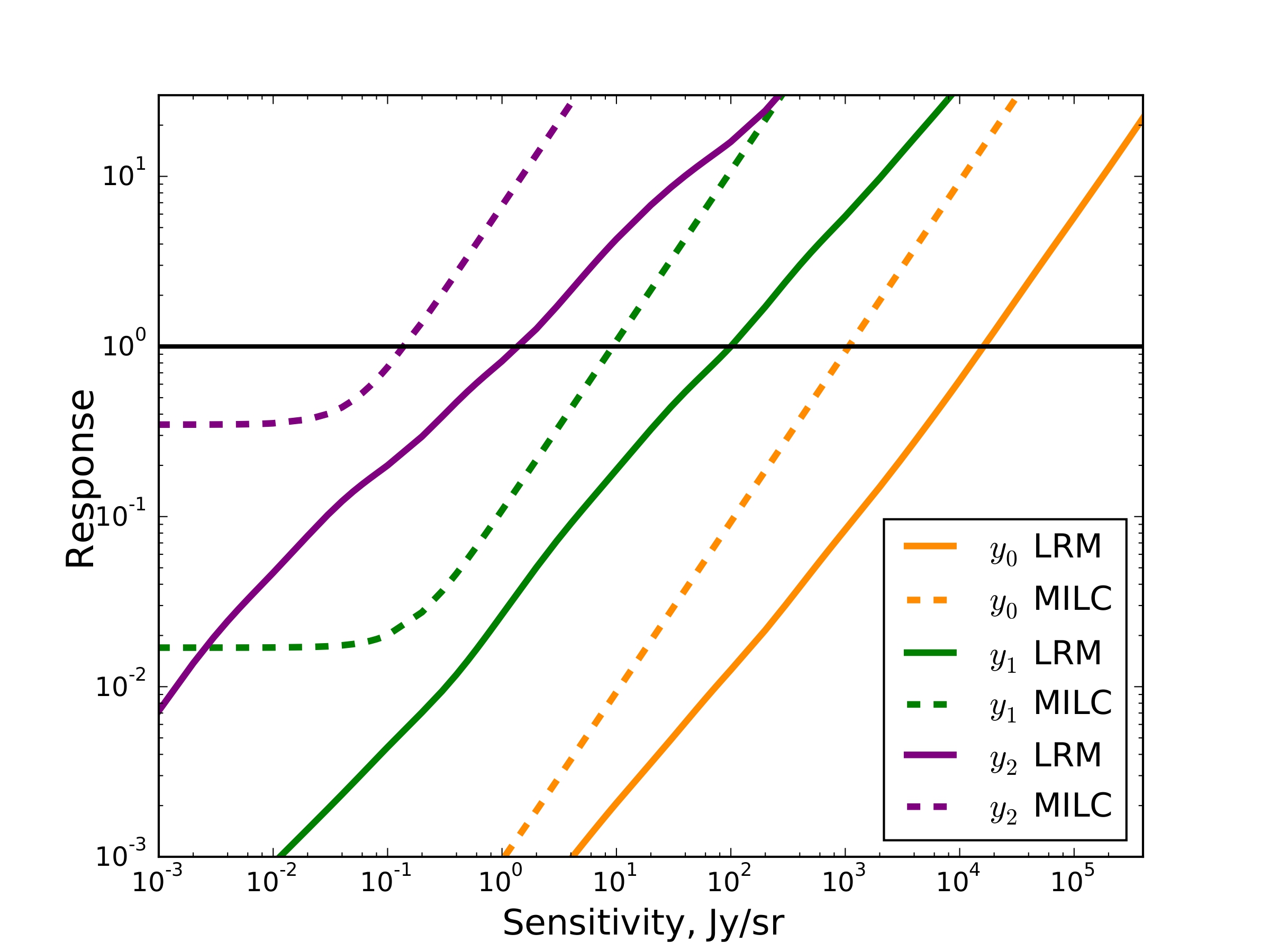}
\includegraphics[width=0.54\textwidth]{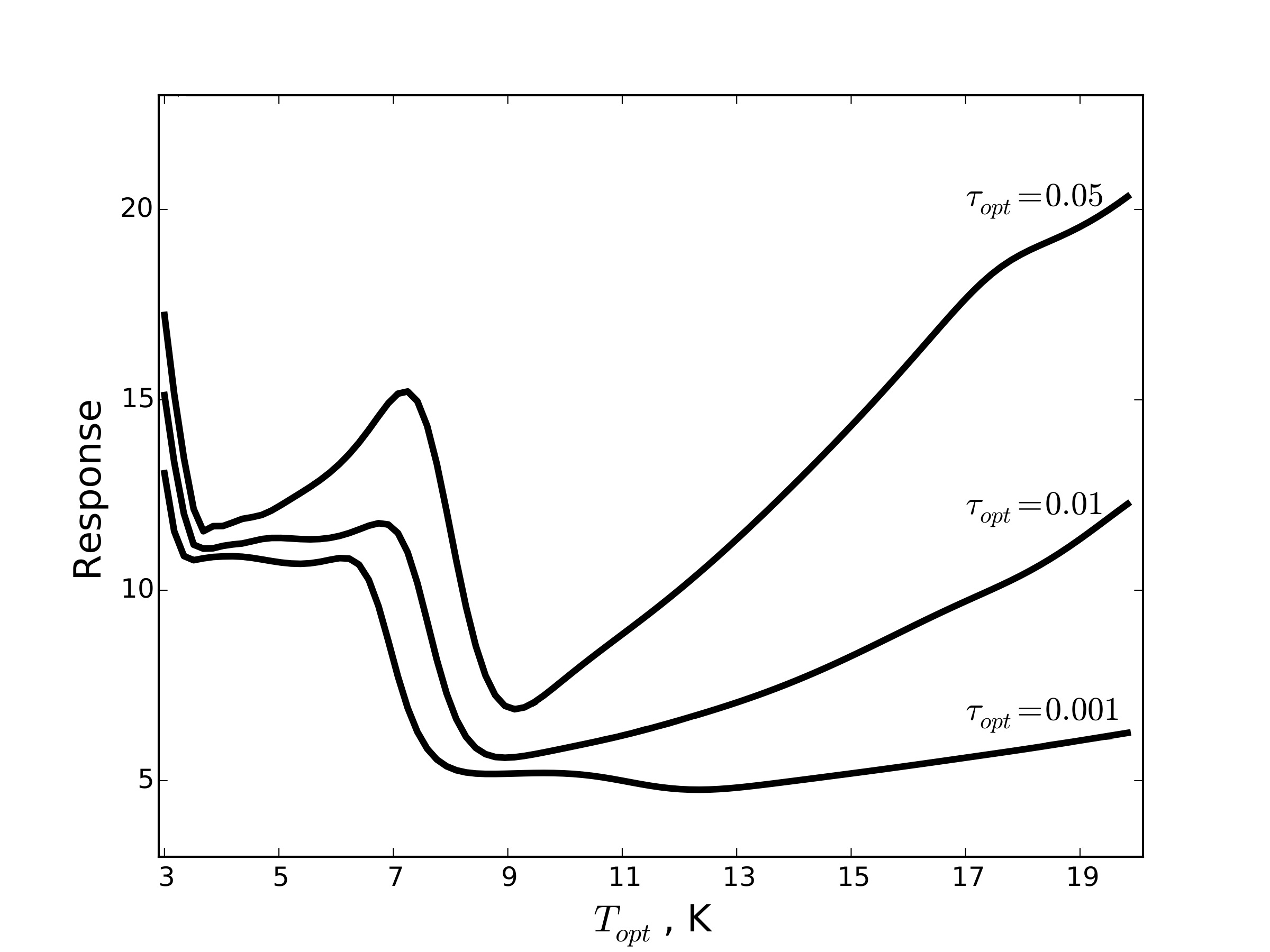}}
\caption{{\it Left panel.} The results of extracting $y$ distortion ($y_0$) and its first two relativistic corrections ($y_1$ and $y_2$) for the LRM and MILC approaches.
{\it Right panel.} The response to  foreground+noise as a function of the temperature of the instrument optics for
different emissivity values.
}
\label{pic:4}
\end{figure*}

We also have managed to obtain a pivotal result regarding the optimal temperature of the telescope mirror for measuring $\mu$ distortion (Fig. (\ref{pic:4})). It was shown~\cite{ma} that
the optimal result is achieved when the temperature of an optical system is close to 9 K.

\section{Conclusion}

CMB spectral distortions contain a wealth of information that is virtually unobtainable by other observational methods. To extract these distortions, we have introduced an algorithm whose main feature is its weak sensitivity to the foreground spectral shapes. To minimize the responses to both noise and foreground, we need to know the domains where parameters of the foregrounds vary and the upper bounds on the foreground amplitudes.

We have compared LRM and MILC methods. The total MILC response to noise significantly exceeds LRM results for any sensitivity value. We also have established an optimal temperature of the telescope mirror ($9$ K in our case) which allows to evade "phantom" instrumental signals that may resemble the shape of the $\mu$ distortion.

\section*{Acknowledgments}
 We would like to thank Jacques Dumarchez and other PASCOS 2024 organizers for an opportunity to present our work. The research was supported by an RSF grant \textnumero\;24-22-00230.

\end{document}